\documentclass[twocolumn, 10pt]{article}

\usepackage{fullpage}
\usepackage{abstract}

\usepackage{xspace}
\usepackage{multirow}
\usepackage{soul}
\usepackage[usenames,dvipsnames]{color}
\usepackage{graphicx}
\usepackage[normalem]{ulem}
\usepackage{array}
\usepackage{appendix}
\usepackage{rotating}
\usepackage{sectsty}
\usepackage{setspace}
\usepackage{subfigure}
\usepackage{lineno}
\usepackage{amsmath}

\def\textTM{\texttrademark\xspace}

\def\dbline     {\noalign{\vskip 0.10truecm\hrule\vskip 0.05truecm\hrule\vskip 0.10truecm}}

\newcommand{\cc}{\ensuremath{\mathrm{\,cc}}\xspace}

\def\vsevfi {\ensuremath{V_{75}}\xspace}
\def\vhund  {\ensuremath{V_{100}}\xspace}
\def\vhuntw {\ensuremath{V_{120}}\xspace} 
\def\vhuntwf{\ensuremath{V_{125}}\xspace} 
\def\vhunfi {\ensuremath{V_{150}}\xspace}

\def\pvhund {\ensuremath{\vhund^   {\mathrm{Prostate}}}\xspace}
\def\pvhunfi{\ensuremath{\vhunfi^  {\mathrm{Prostate}}}\xspace}

\def\uvhuntw {\ensuremath{\vhuntw^ {\mathrm{Urethra}}}\xspace}
\def\uvhuntwf{\ensuremath{\vhuntwf^{\mathrm{Urethra}}}\xspace}
\def\uvhunfi {\ensuremath{\vhunfi^ {\mathrm{Urethra}}}\xspace}

\def\bvsevfi{\ensuremath{\vsevfi^{\mathrm{Bladder}}}\xspace}

\def\rvsevfi{\ensuremath{\vsevfi^{\mathrm{Rectum}}}\xspace}

\def\hypoNull{\ensuremath{\mathrm{H}_0}\xspace}

\def\IPSA{IPSA\textTM }
\definecolor{lightgray}{gray}{0.50}

\title{Dosimetric equivalence of non-standard HDR brachytherapy catheter patterns}
\author{J.~A.~M. Cunha, I-C. Hsu, and J. Pouliot \\
Department of Radiation Oncology \\
University of California\\
San Francisco, CA, USA\\
{\small CunhaA@RadOnc.UCSF.Edu}\\
}
\date{12 November 2008}

\begin{document}
\hyphenation{brachy-ther-a-py}
\twocolumn[ 
\maketitle 
\vspace{-0.2in}
\begin{onecolabstract} 

{\bf Purpose:} To determine whether alternative high dose rate prostate brachytherapy catheter patterns can result in similar or improved dose distributions while providing better access and reducing trauma.  

{\bf   Materials and Methods:} Standard prostate cancer high dose rate brachytherapy uses a regular grid of parallel needle positions to guide the catheter insertion.  This geometry does not easily allow the physician to avoid piercing the critical structures near the penile bulb nor does it provide position flexibility in the case of pubic arch interference.  This study used CT datasets with 3 mm slice spacing from ten previously-treated patients and digitized new catheters following three hypothetical catheter patterns: conical, bi-conical, and fireworks.  The conical patterns were used to accommodate a robotic delivery using a single entry point.  The bi-conical and fireworks patterns were specifically designed to avoid the critical structures near the penile bulb.  For each catheter distribution, a plan was optimized with the inverse planning algorithm, \IPSA, and compared with the plan used for treatment. Irrelevant of catheter geometry, a plan must fulfill the RTOG-0321 dose criteria for target dose coverage ($\pvhund>90\%$) and organ-at-risk dose sparing ($\bvsevfi<1$cc, $\rvsevfi<1$cc, $\uvhuntwf<<1$cc). 

{\bf Results:} The three non-standard catheter patterns used 16 non-parallel, straight divergent catheters, with entry points in the perineum.  Thirty plans from ten patients with prostate sizes ranging from 26 to 89 cc were optimized.  All non-standard patterns fulfilled the RTOG criteria when the clinical plan did. In some cases, the dose distribution was improved by better sparing the organs-at-risk.

{\bf Conclusion:} Alternative catheter patterns can provide the physician with additional ways to treat patients previously considered unsuited for brachytherapy treatment (pubic arch interference) and facilitate robotic guidance of catheter insertion. In addition, alternative catheter patterns may decrease toxicity by avoidance of the critical structures near the penile bulb while still fulfilling the RTOG criteria.\newline


\end{onecolabstract} 
\vspace{0.2in}
]

\section{Introduction}
Excellent survival rates and local control have been reported using prostate high dose rate (HDR) brachytherapy~\cite{Grills_2004-CU, Demanes_2005-IJROBP, Hsu_2005-B,   Yoshioka_2006-RO}.  The next challenge facing the field involves reducing trauma and side effects, and removing the obstacles that prevent patients from being qualified for brachytherapy treatment. Trauma to the critical structures near the penile bulb (urethra, nerves, vessels, and penile bulb) from the insertion of the brachytherapy catheters may be a key component of the former, while pubic arch interference contributes to the latter~\cite{Strang_2001-RJ,   Henderson_2003-BJ}.  It has been shown that radiation delivered to the penile bulb can lead to erectile dysfunction~\cite{Fisch_2001-U,   Merrick_2001-IJROBP, Vargas_2005-JU}.  This may be secondary, though, to damage to the cavernous arteries running along the side of the penile bulb where physical trauma has been shown to cause erectile dysfunction~\cite{Munarriz_1995-JU}. In light of these results, it is reasonable to hypothesize that trauma from needle puncture to the critical structures near the penile bulb (CSNB) may be deleterious to erectile function\footnote{Where applicable, the CSNB is inclusive of the penile bulb.}.  Both pubic arch interference and trauma to the CSNB from needle puncture are direct consequences of the standard setup used for the transperineal insertion of the brachytherapy catheters that allows for little deviation from a parallel implantation geometry.

Since the advent of anatomy-based inverse planning early in this decade, there has been discussion in the field about whether the standard template-based brachytherapy procedure is necessary or whether it is viable to use alternative geometries.  For example, inverse planning has allowed for the development of template-less implantation of HDR brachytherapy catheters~\cite{Kim_2007-JACMP}, though in this case the implanted catheters are roughly parallel, each with a different skin-entry point in the perineum.  The beginning of this decade also evidenced pioneering work in understanding the feasibility of alternative catheter patterns in the context of Prostate Permanent-seed Implant brachytherapy (PPI).  Van Gellekom et al. were the first to introduce the concept of using a single needle to implant PPI seeds~\cite{vanGellekom_2004-RO}.  This was inspired by the desire to use a robot for seed placement and resulted in a conical seed geometry.  Fu et al.  followed with work that applied inverse planning techniques to a conical PPI implant geometry~\cite{Fu_2005-BJ, Fu_2006-MP}; and Van den Bosch et al. modified our \IPSA~\cite{Lessard_2001-MP, Lessard_2006-MP,   Pouliot_2005-Chapter} code to perform dose coverage studies of conical and bi-conical PPI seed geometries~\cite{vandenBosch_2008-RO}.

The work discussed here moves away from PPI and explores the concept of alternative catheter patterns in the context of HDR brachytherapy. We focus on determining the clinical feasibility of alternative HDR brachytherapy catheter patterns.  To this end we do not modify the optimization code, \IPSA, in any way.  For the purpose of this paper, clinical feasibility is defined in the physics-based terms of determining whether clinically acceptable dose distributions can be generated from non-standard catheter patterns.  

Standard clinical HDR brachytherapy practice uses a mechanical template of a regular geometry to guide linear catheters transperineally into the body in parallel with little or no curvature.  This setup nearly guarantees that the catheters pierce the CSNB and has little flexibility to work around a pubic arch that extends inferior of the prostate.  We relaxed the constraint on parallelism to determine the dosimetric quality achievable via HDR brachytherapy using non-parallel catheter arrangements.  The three geometries presented show that non-standard catheter patterns: (1) are able to circumvent pubic arch interference; (2) are able to avoid piercing the CSNB; and (3) accommodate a single-point-of-entry robotic system.

\section{Materials and Methods}
\subsection{Planning system and dataset}
All catheter digitization and DVH generation was performed on Nucletron's clinical brachytherapy platform, PLATO. The \IPSA inverse planning optimization engine available through PLATO was used to generate the plans. CT datasets from ten previously-treated patients were used and catheters were digitized in three hypothetical patterns: conical, bi-conical, and fireworks.  Since 16 catheters were used for all the actual treatments, we restricted the new catheter patterns to no more or less than 16 catheters.  Treatment plans were generated with a prescription dose of 950 cGy per fraction.

\subsection{Catheter patterns}
\begin{figure*}[thpb]
 \centering
 \includegraphics[width=\textwidth]{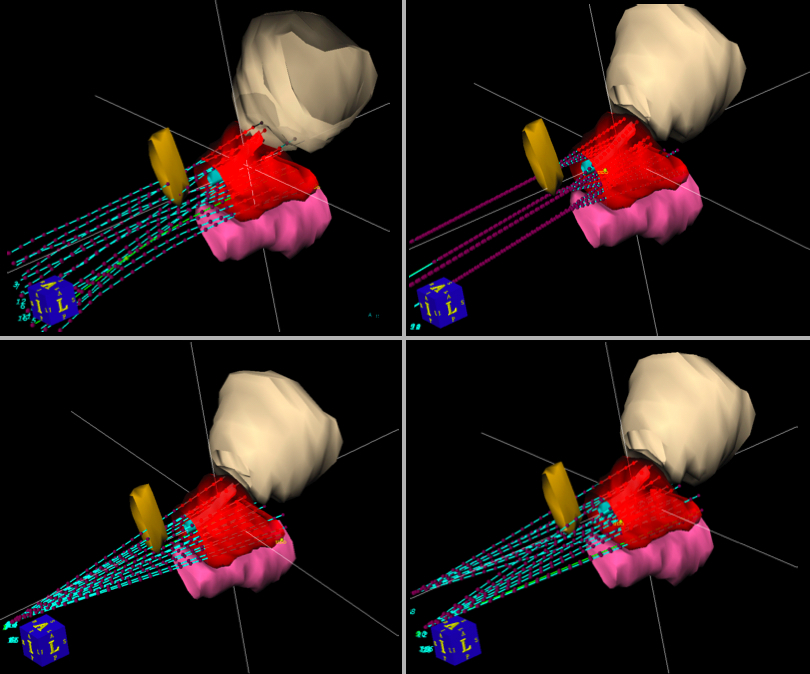} 
 \caption{A computer generated image of the catheter arrangements used in    this study.  All four images are for the same patient. The bladder is in the upper right, the prostate (target volume) in the center and contains the urethra, and the rectum on the bottom.    The contoured region to the left of the prostate represents both the penile bulb and the CSNB.  In the original dose planning, it was contoured as a general region of avoidance. The catheters enter from the lower left of the figure, which corresponds to the patient's perineum.  \emph{Upper left:} The standard implant used in the actual patient treatment.  Note that the catheters pierce the  contoured bulb/CSNB.    \emph{Upper right:} A hypothetical Fireworks distribution.  The Fireworks configuration of catheters allows for avoidance of the contoured bulb/CSNB.    \emph{Lower left:} The conical shape of the implant allows for a single entry point for all catheters. The apex of the cone is located at the patient's perineum; catheters spread as    they approach the prostate to cover the prostate while avoiding the urethra.  \emph{Lower right:}    The bi-conical catheter pattern can accommodate a single entry point, circumvent the contoured bulb/CSNB, and access parts of the target volume behind the urethra inaccessible via the standard or conical approach.}
 \label{fig:CatheterPatterns}
\end{figure*}

\begin{figure*}[thpb]
 \centering
 \includegraphics[width=\textwidth]{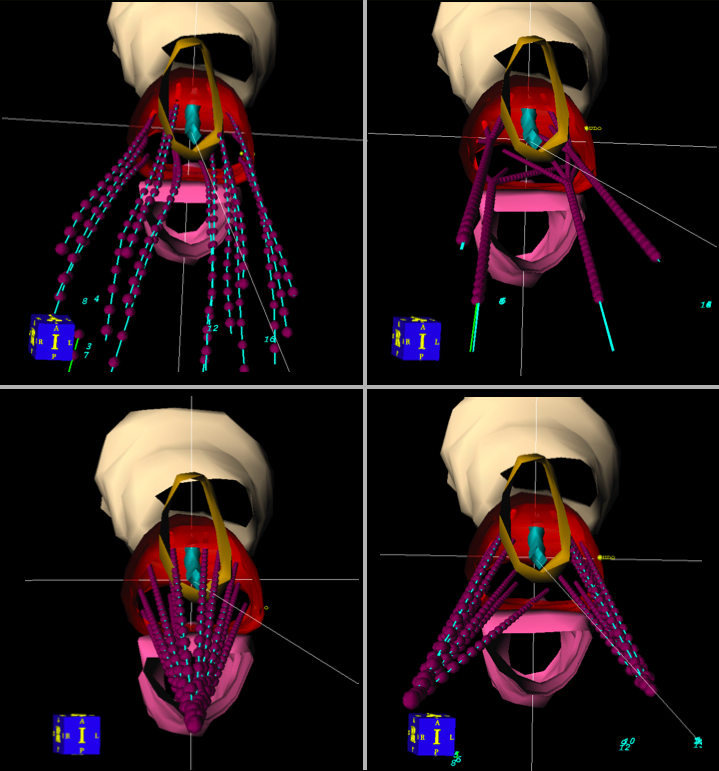}
 \caption{{\bf Cunha: Dosimetric equivalence of non-standard catheter
     patterns.}  Color figure available online.  The transverse view
   of the same images presented in Figure~\ref{fig:CatheterPatterns}.
   Here the bladder is in the upper portion of the image, followed by
   the contoured bulb/CSNB, the prostate, the
   urethra, and finally the rectum in the lower portion. The catheters
   enter from the lower foreground (the patient's perineum) and
   proceed into the page.  \emph{Upper left}: A standard implant used
   in the actual patient treatment.  \emph{Upper right}: The same
   patient, but with a hypothetical Fireworks distribution.
   \emph{Lower left}: A single-cone catheter pattern. \emph{Lower right}: A bi-conical catheter pattern. }
 \label{fig:CatheterPatterns-supinf}
\end{figure*}

\begin{figure*}
 \begin{center}
 \includegraphics[width=\textwidth]{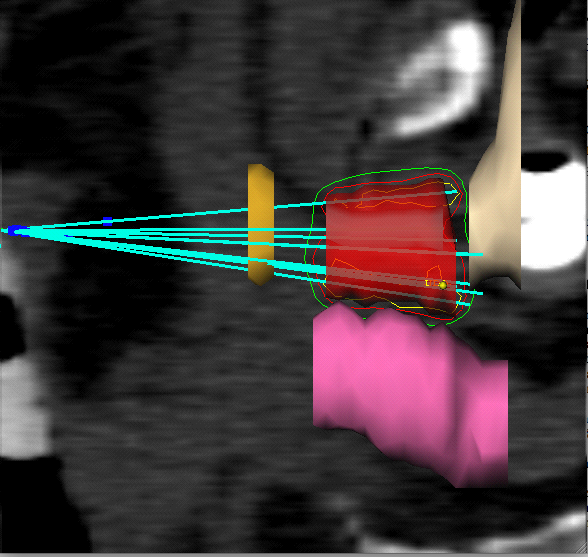}
 \caption{ A sagittal view of one of the patients in our study that shows the conical pattern used. On the far right (in white) is the balloon at the end of the Foley Catheter inside the bladder.  The inferior slices of the bladder are contoured. To the left of the bladder is the prostate surrounding the urethra.  Below the prostate is the rectum and above it is the pubic arch.  To the left of the prostate is the contoured bulb/CSNB.  The catheters enter from the perineum on the left of the image.}

 \label{fig:sagittal}
 \end{center}
\end{figure*}

The conical and bi-conical geometries were inspired by future robotic delivery systems that may require a single point of entry for the catheters~\cite{Stoianovici_2007-MITAT, DiMaio_2007-CAS}.  The bi-conical and fireworks patterns were specifically designed to avoid the CSNB.  The contoured penile bulb was used to represent both the penile bulb itself and the harder-to-image CSNB. The three patterns examined in the study are shown in Figures~\ref{fig:CatheterPatterns} and~\ref{fig:CatheterPatterns-supinf}.  Along with a conical catheter pattern, the 3D anatomy reconstruction is overlaid on a sagittal 2D image slice in Figure~\ref{fig:sagittal}.  For this study, all catheters were digitized as straight needles with no curvature.

\subsubsection{Conical patterns}
The conical catheter patterns examined in this study have their apex situated in the first subcutaneous CT slice of the patient image set. The effect of placing the apex one slice extra in both the super- and subcutaneous directions was examined and found to be negligible.

To create the single-cone catheter pattern, the digitization process proceeded using the actual implant digitization as a guide.  The base of every catheter was digitized in the perineum, all at a single point. Next, all other digitized catheter positions were removed, preserving only the most superior digitized point for each catheter (i.e., the tip of the catheter).  The result is a conical pattern of straight catheters of various lengths depending on the depth of the prostate and bladder. Unlike in~\cite{vandenBosch_2008-RO}, all catheters passed entirely through the prostate and none needed to be terminated early due to urethral interference. None of the newly digitized catheters passed through the urethra, rectum, or bladder.

Digitization of the two-cone catheter patterns started with the single-cone pattern described above.  Maintaining the catheter tip location, the bases of the catheters were separated into two groups.  Each group was collected at its respective entry point, one on each side of the perineum to from two cone-shaped patterns.  Because one of the goals of this study was to evaluate the ability to avoid puncture of the CSNB while maintaining adequate dose coverage, the CSNB were avoided for all catheters.

\subsubsection{Fireworks catheter pattern}
In this case the physician would insert four larger-than-normal-gauge catheters into the perineum and past the CSNB, creating a conduit for the standard source-bearing catheters, which may be inserted without each having to each pierce the skin. At the top of the conduit, the source-bearing catheters diverge to access all parts of the prostate.

These patterns were again based on the catheter placement used in the actual treatment.  For this step, the tips of the catheters were left stationary, but all other digitized locations were removed. The four larger-gauge catheters---one for each quadrant of the prostate as viewed transversely---were digitized from the perineum to the last slice with a contoured CSNB.  The treatment catheters were then split into four four-catheter groups and connected to the guide catheters.

A fireworks-like pattern would allow for complete physical avoidance of the CSNB while still maintaining a quality dose distribution and was inspired by the fact that impotence is a common side affect of brachytherapy not only because of the radiation delivered to the penile bulb but also from the physical puncturing of the CSNB by the brachytherapy catheters.

\section{Results}
Table~\ref{tab:DosiVars} shows the results from this study. The standard criteria recommended by the RTOG-0321 protocol~\cite{RTOG0321} are: $\pvhund>90\%$, $\bvsevfi<1\cc$, $\rvsevfi<1\cc$, and $\uvhuntwf<<1\cc$. For each of the ten patients, the RTOG-0321 criteria are presented. In our clinic, \IPSA consistently allows us to obtain $\uvhuntwf=0\cc$, therefore we use (and report here) the more-strict \uvhuntw.  (Heretofore, when the RTOG-0321 protocol is mentioned in the context of this study we mean to reference \uvhuntw rather than \uvhuntwf.) We also include \uvhunfi, \pvhunfi, and the prostate homogeneity index, HI = $\frac{\pvhund -\pvhunfi} {\pvhund}$ since these are used in our clinic to augment the RTOG-0321 criteria. The HI is desired to be greater than 0.6.  The plan used for the actual patient treatment is listed in the column marked ``Original'' and was considered the control sample to which the non-parallel-catheter plans can be compared.  All three of the experimental catheter configurations were able to satisfy the RTOG-0321 criteria when the clinically used plan did.  Only two of the 30 plans had a HI less than 0.60 (0.59 in both cases).

The two cases in which the catheter patterns (both the actual implant and the experimental patterns) were not able to fulfill the RTOG-0321 criteria were exceptional anatomical cases.  Case 8 had a long thin superior extension of the prostate posterior to the bladder that was hard to accommodate without a suboptimal dose delivered to the bladder. The prostate size in Case 9 was large (89.0 cc) and would have benefited from the use of more catheters. We did not increase the catheter number in order to preserve this variable as constant across all studies for this patient.  Despite the large contoured penile  bulb in this case (and thus large region of possible location for the CSNB), the CSNB-sparing patterns were able to obtain a dose coverage of similar quality to the actual treatment while still avoiding puncture of the bulb/CSNB region\footnote{Note that our clinic uses 16 catheters for every HDR brachytherapy plan.  This is because our experience has been that the benefits of increased radiological coverage are offset by the side effects which result from normal tissue trauma when more than 16 catheters are physically inserted into the body.  It was our desire to to minimize the work flow changes needed to implement this method.  As such, we kept the number of catheters the same as would be used in our actual clinical setting.}. Prostate size, however, is not necessarily a limiting factor---Case 4 has clinically acceptable results even with a prostate volume of 80.2 cc.

\begin{figure*}
 \begin{center}
 \includegraphics[width=\textwidth]{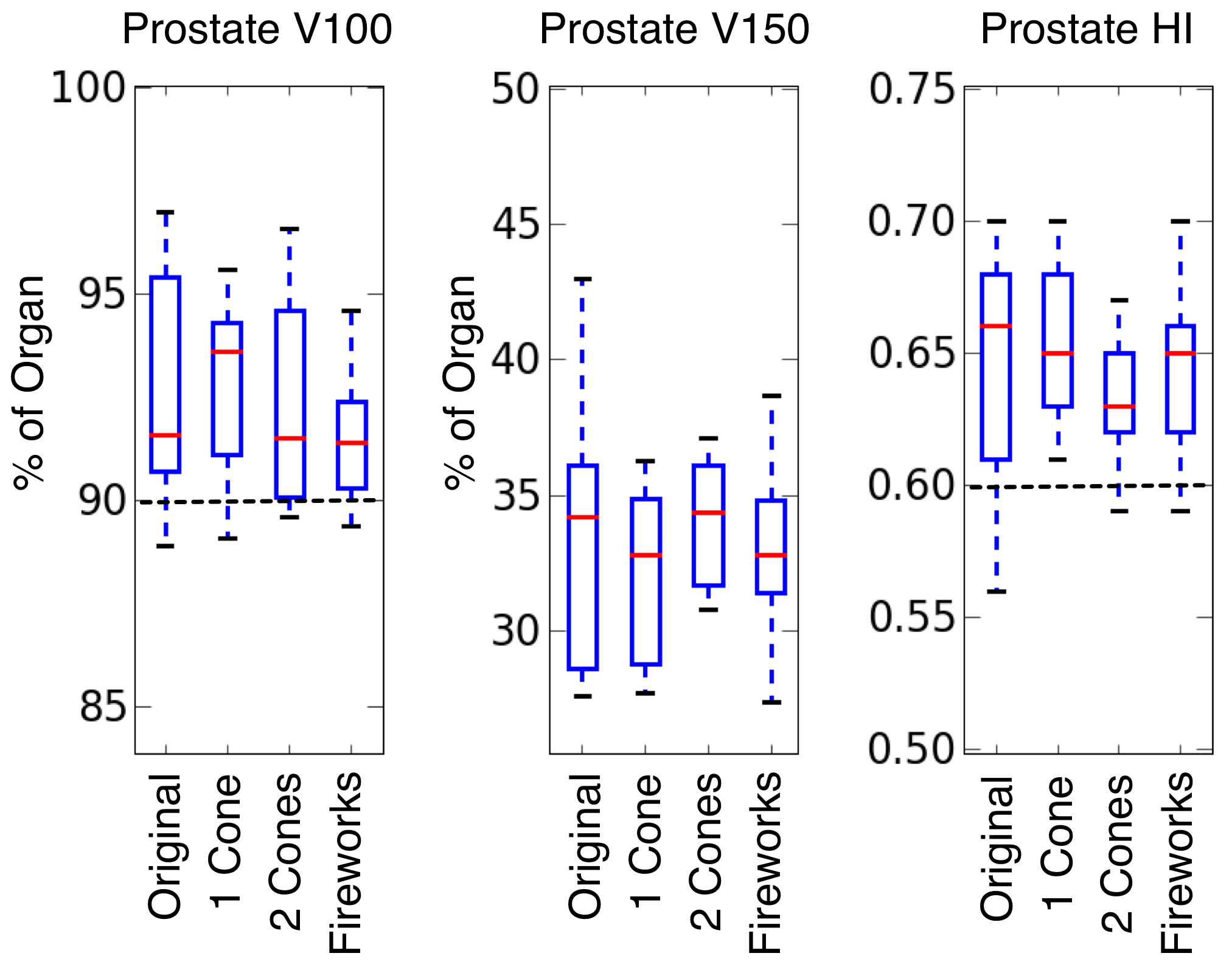}
 \caption{\small Box-and-whisker plots of the data presented in    Table~\ref{tab:DosiVars} for prostate. HI is the prostate dose homogeneity index.    A standard median-value box-and-whisker construction is used: the box encompasses    the inner two quartiles (the range of which is the inner quartile range, IQR)    with a hash at the median value of the distribution; the whiskers end at the    lowest(highest) measured value within the box edge minus(plus) $1.5\times    \mathrm{IQR}$. If there is no value within this range the whisker is not drawn    (i.e. the whisker ends at the edge of the box). The key element to note here is    similarity of all results.  The RTOG-0321 recommendation of 90\% for the \pvhund    and our clinic's recommendation of the prostate HI$>0.6$ are indicated by the    dashed horizontal line.  Friedman's Statistics, $S$, which compare the set of four studies (control, conical, bi-conical, and fireworks) are $S(\pvhund)=2.19$ and $S(\mathrm{HI})=3.18$. These values are both less than the cutoff of 7.81 for an alpha of 0.05. We are therefore at least 95\% confident that any differences in the plans is due purely to random statistical fluctuations.}

 \label{fig:boxTarget}
 \end{center}
\end{figure*}

\begin{figure*}
 \begin{center}
 \includegraphics[width=\textwidth]{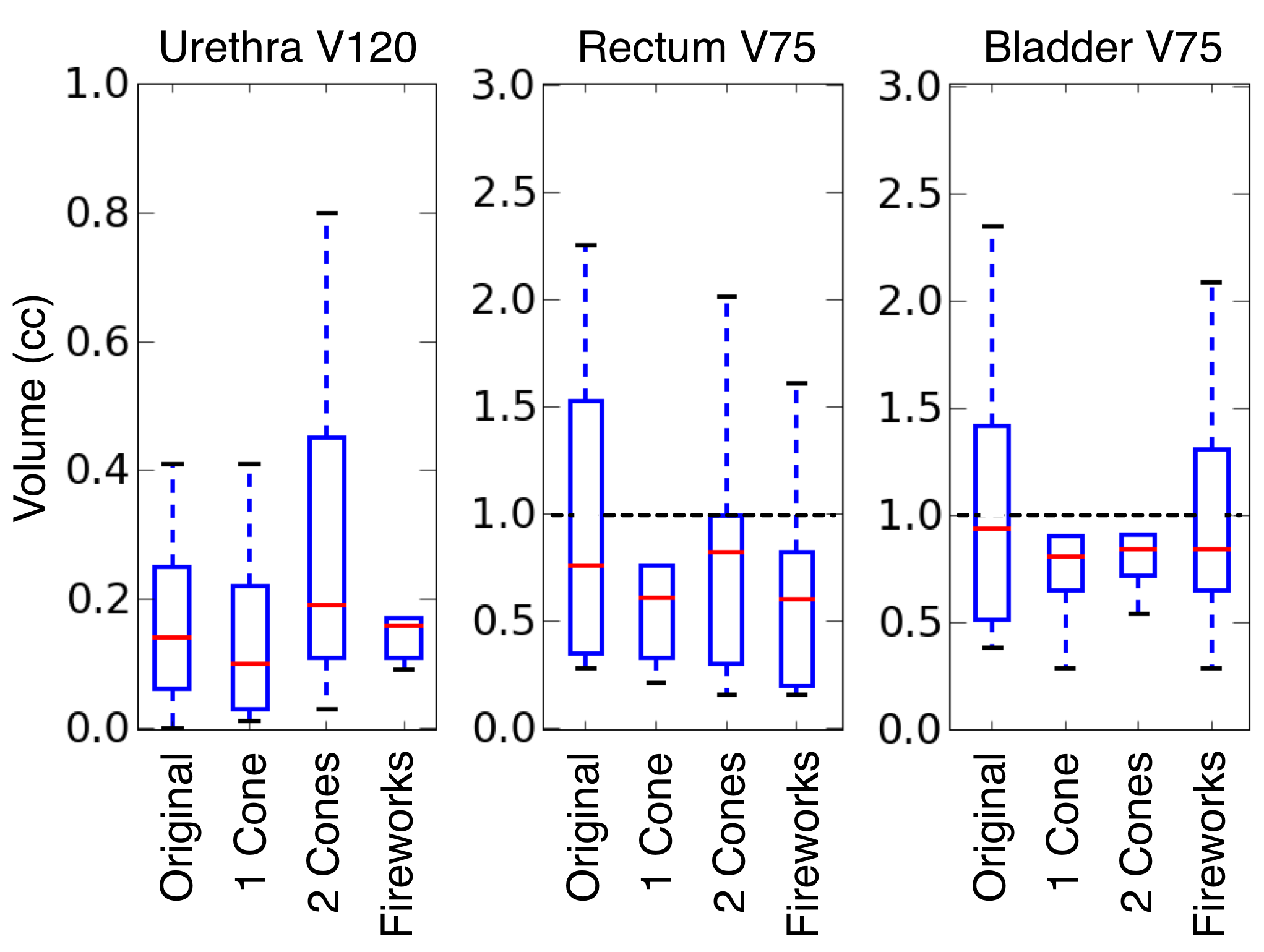}
 \caption{\small Box-and-whisker plots of
 the data presented in Table~\ref{tab:DosiVars} for the organs at
 risk. The box and whisker generation is the same as for
 Figure~\ref{fig:boxTarget}.  (The lack of whiskers on some of the
 plots is directly related to the smallness of the box, i.e. it is a
 result of the increase in consistency of that dosimetric variable
 across all patients.) As in Figure~\ref{fig:boxTarget}, the critical
 point here is to note the similarity of the alternative patterns with
 respect to the Original.  This is critical given that the alternative
 patterns are just as good as the Original, parallel-catheter plans {\em
 and} they provide for avoidance of the contoured bulb/CSNB or pubic
 arch. (Note, there is a change of scale versus the urethra plot
 for those of the bladder and rectum.)  Our clinical requirement of
 $\uvhuntw<1.0$ cc is indicated by the limit of the ordinate in the
 Urethra plot.  The RTOG-0321 recommended limit of $\vsevfi<1.0$ cc is
 indicated as the dashed horizontal line for the latter two.
 Friedman's statistics, $S$, which compare the set of four studies
 (control, conical, bi-conical, and fireworks) are $S(\uvhuntw)=3.63$,
 $S(\rvsevfi)=5.88$, $S(\bvsevfi)=2.97$. These values are all less
 than the cutoff of 7.81 for an alpha of 0.05. We are therefore at
 least 95\% confident that any differences in the plans is due purely
 to random statistical fluctuations.}

 \label{fig:boxOAR} 
 \end{center}
\end{figure*}

Box-and-whisker diagrams of the \pvhund, \pvhunfi, HI, \uvhuntw, \rvsevfi, and \bvsevfi data from Table~\ref{tab:DosiVars} are shown in Figures~\ref{fig:boxTarget} and~\ref{fig:boxOAR}.  For each dosimetric variable, the distribution of values for each catheter pattern is shown. A standard construction of rank-based box-and-whisker plots was used: the box encompasses the inner two quartiles (the range of which is the inner quartile range, IQR) with a hash at the median value of the distribution; the whiskers end at the lowest(highest) measured value within the box edge minus(plus) $1.5\times \mathrm{IQR}$. If there is no value within this range the whisker is not drawn (i.e. the whisker ends at the edge of the box).

It is clear from Figure~\ref{fig:boxTarget} that the ability to deliver a clinically acceptable dose to the prostate target is independent of the catheter pattern used.  In addition, the distribution of doses in Figure~\ref{fig:boxOAR} shows that for the surrounding organs, there is no clinically significant difference between the alternative catheter patterns and the original, clinically-used plan.

The measured dosimetric variables are not expected to follow a Gaussian distribution (or any other parametric distribution).  Therefore, the quantitative statistical significance of our results was determined via a Friedman test. This non-parametric rank test determines whether the results across all tests (original, conical, bi-conical, and fireworks) are consistent.  Consistency is defined by evaluating whether the null hypothesis, \hypoNull (all treatments have identical results) is valid. The alternative hypothesis is that at least one treatment is different from at least one other treatment. The validation of \hypoNull is quantified by the test statistic, $S$,
\begin{equation}
S = \frac{12}{np(p+1)}\left(\sum_{j=1}^p(R_j^2)\right) - 3n(p+1) , 
\end{equation}
where $R_j$ is sum of the ranks for catheter pattern $j$, $n$ is the number of patients, and $p$ is the number of catheter patterns studied~\cite{Friedman_1937-JASA}.  

For this experiment, $n=10$ and $p=4$. The degrees of freedom is given by $p-1$. For $n=10$, three degrees of freedom, and a significance level of 5\% ($\alpha = 0.05$), the cutoff value for the test statistic is $S=7.81$.  If \hypoNull is true and this experiment is run an infinite number of times, 95\% of the results will yield a value $S\leq7.81$ and 5\% of the results will yield a value of $S>7.81$.

We obtain the following $S$ values: 
\begin{align*}
 S(\pvhund) = 2.19 \\
 S(\mathrm{HI}) = 3.18 \\
 S(\uvhuntw)=3.63 \\
 S(\rvsevfi)=5.88 \\
 S(\bvsevfi)=2.97
\end{align*}
Since all values are less than 7.81, we are at least 95\% certain that the differences we see are entirely due to random fluctuations and therefore {\em the null hypothesis is valid: There is no statistically significant difference   between any of the alternative plans and the original.}

\section{Discussion}
There has been discussion in the literature about conical needle configurations~\cite{vanGellekom_2004-RO, Fu_2005-BJ,   Fu_2006-MP, vandenBosch_2008-RO}, but all are in the context of PPI brachytherapy.  The work presented in Reference~\cite{vandenBosch_2008-RO}, for example, modified the \IPSA code to examine the potential of conical catheter patterns.  However, the work presented here addresses the viability of conical catheter patterns in HDR brachytherapy and examines the feasibility of non-standard catheter patterns using clinically-available software.

To reiterate the focus of this study, we have shown that while standard HDR brachytherapy uses a regular-grid template for catheter placement, that particular configuration is not necessary to achieve a clinically acceptable plan. Parallel-catheter-based brachytherapy techniques were developed to help ensure a uniform conformal dose distribution prior to the advent of inverse planning and present-day imaging technology.  However, catheter implant technology is moving past the need for a fixed rectangular template.  In fact, we have been using a freehand TRUS-guided system that incorporates non-parallel needles since 1997~\cite{Kim_2007-JACMP}.  The catheter implantation still results in mostly parallel catheters (though this is not a requirement).

It should also be noted that robotic delivery devices that will be coming to market in the near future~\cite{Stoianovici_2007-MITAT, DiMaio_2007-CAS} have the potential to provide placement on the order of millimeters. In this context, the use of a template is no longer a necessity; and, alternative catheter placement patterns should be explored so that the best patterns can be exploited.  For example, because of their location on the lateral portions of the prostate, a conical catheter pattern with its apex centrally located could allow for easier avoidance of the neurovascular bundles.

Data regarding long-term efficacy of HDR brachytherapy are continuing to emerge. For patients who are suited for HDR brachytherapy treatment, a number of studies have been released in recent years.  In 2004, William Beaumont Hospital reported biochemical control in 98\% of patients at a followup of 35 months~\cite{Grills_2004-CU}.  In 2005, with a median followup of 7.25 years, the California Endocurietherapy Cancer Center reported a general clinical control rate of 90\% with a cause-specific survival rate of 97\%~\cite{Demanes_2005-IJROBP}.  In 2005, UCSF reported 4-year overall and disease-free survival rates of 95\% and 92\%, respectively, for patients treated with an HDR boost to external beam radiotherapy~\cite{Hsu_2005-B}.  And in 2006, Osaka University reported 3- and 5-year local control rates of 100\% and 97\%~\cite{Yoshioka_2006-RO}. However, there is still much progress to be made in reducing side effects of treatments and increasing the percentage of patients for whom brachytherapy is an option. Moving away from a template-based HDR brachytherapy procedure is a key way to move forward. We have shown here that alternative catheter patterns can result in clinically identical dose coverage and therefore equal survival rates. The main argument here, however, is not that clinically acceptable HDR brachytherapy dose profiles can be achieved, but rather that, just as for PPI procedures, it can be done \emph{while avoiding penetration of the  CSNB and/or avoiding pubic arch interference}.  

Note that while we explicitly avoided piercing the CSNB in the bi-conical and fireworks catheter patterns, no attempt was made to further reduce the dose to the CSNB.  It has been shown~\cite{Pouliot_2005-Chapter} that inverse planning can reduce the dose to the penile bulb at little or no cost to the target coverage or sparing of the other organs at risk.  It may also be possible to generate plans with these new catheter patterns using a smaller number of catheters, thus possibly further reducing trauma.

\section{Conclusion}
We have examined the feasibility of using alternative HDR brachytherapy catheter patterns to produce clinically-acceptable dose plans for prostate cancer treatment.  Using ten previously-treated patient anatomical data and three alternative catheter geometries---conical, bi-conical, and fireworks---we have shown that it is possible to generate dose plans that are clinically acceptable and at least as good as the plan used in the actual patient treatment. We have changed the optimization procedure as little as possible from our clinic's standard work flow in order to show that minimal fine tuning is necessary and therefore implementation in the clinic, in regards to dose planning, is trivial. 

\section*{Acknowledgement}
This work was supported by Nucletron.

\begin{table*}
 \caption{ 
 Dosimetric plan-evaluation criteria for the actual treatment plan
 (Original) and the three other plan geometries examined in this
 study. All alternative catheter patterns satisfy the RTOG-0321 protocol
 criteria when the original plan did. Note that in most cases, the
 alternative geometries are either as good or better than the clinical
 plan.}

 \label{tab:DosiVars} 
 \begin{center} 
  {\footnotesize
   \begin{tabular}{l|cccc||l|cccc} 
    \dbline 
    Case 1 (V=42.3cc)    & Original & Cone & Cone2 & Firework  &  Case 6 (V=33.3cc)   & Original & Cone & Cone2 & Firework \\
 Bladder \vsevfi (cc)& 1.42     & 0.29 & 0.91  & 1.31      &  Bladder \vsevfi (cc)& 0.45     & 0.66 & 0.82  & 0.30 \\
 Rectum  \vsevfi (cc)& 0.52     & 0.59 & 0.99  & 0.82      &  Rectum  \vsevfi (cc)& 0.76     & 0.65 & 0.81  & 0.63 \\
 Urethra \vhund (cc) & 0.97     & 0.85 & 1.02  & 0.95      &  Urethra \vhund (cc) & 0.95     & 0.96 & 0.85  & 0.94 \\
 Urethra \vhuntw (cc)& 0.30     & 0.03 & 0.80  & 0.46      &  Urethra \vhuntw (cc)& 0.18     & 0.10 & 0.03  & 0.10 \\
 Target \vhund (\%)  & 88.9     & 89.1 & 89.6  & 89.4      &  Target \vhund (\%)  & 90.7     & 93.6 & 91.5  & 90.3 \\
 Target \vhunfi (\%) & 28.2     & 31.4 & 32.3  & 34.8      &  Target \vhunfi (\%) & 28.6     & 27.7 & 34.8  & 29.7 \\
 Target HI           & 0.58     & 0.65 & 0.64  & 0.61      &  Target HI           & 0.69     & 0.70 & 0.62  & 0.67 \\
                     &          &      &       &           &                      &          &      &       &         \\
Case 2 (V=44.5cc)    & Original & Cone & Cone2 & Firework  &  Case 7 (V=26.4cc)   & Original & Cone & Cone2 & Firework \\
 Bladder \vsevfi (cc)& 0.98     & 0.81 & 0.72  & 0.66      &  Bladder \vsevfi (cc)& 0.77     & 0.76 & 0.84  & 0.65 \\
 Rectum  \vsevfi (cc)& 0.77     & 0.21 & 0.16  & 0.20      &  Rectum  \vsevfi (cc)& 0.28     & 0.33 & 0.30  & 0.20 \\
 Urethra \vhund (cc) & 1.37     & 1.42 & 1.37  & 1.39      &  Urethra \vhund (cc) & 1.54     & 1.48 & 1.57  & 1.54 \\
 Urethra \vhuntw (cc)& 0.25     & 0.41 & 0.45  & 0.29      &  Urethra \vhuntw (cc)& 0.41     & 0.53 & 0.04  & 0.16 \\
 Target \vhund (\%)  & 97.0     & 94.6 & 94.6  & 94.6      &  Target \vhund (\%)  & 91.6     & 93.5 & 94.2  & 92.4 \\
 Target \vhunfi (\%) & 0.40     & 30.5 & 30.8  & 32.3      &  Target \vhunfi (\%) & 36.1     & 36.3 & 31.1  & 34.6 \\
 Target HI           & 0.58     & 0.68 & 0.67  & 0.66      &  Target HI           & 0.61     & 0.61 & 0.67  & 0.63 \\
                     &          &      &       &           &                      &          &      &       &         \\
Case 3 (V=49.4cc)    & Original & Cone & Cone2 & Firework  &  Case 8 (V=44.5cc)   & Original & Cone & Cone2 & Firework \\
 Bladder \vsevfi (cc)& 0.94     & 0.90 & 0.81  & 0.84      &  Bladder \vsevfi (cc)& 2.35     & 2.39 & 2.24  & 2.09 \\
 Rectum  \vsevfi (cc)& 0.63     & 0.61 & 0.82  & 0.25      &  Rectum  \vsevfi (cc)& 2.25     & 1.67 & 2.01  & 1.47 \\
 Urethra \vhund (cc) & 0.56     & 0.56 & 0.58  & 0.58      &  Urethra \vhund (cc) & 0.93     & 0.97 & 0.94  & 0.96 \\
 Urethra \vhuntw (cc)& 0.11     & 0.03 & 0.13  & 0.17      &  Urethra \vhuntw (cc)& 0.06     & 0.06 & 0.24  & 0.09 \\
 Target \vhund (\%)  & 95.4     & 94.2 & 96.6  & 92.1      &  Target \vhund (\%)  & 91.4     & 91.0 & 90.0  & 91.4 \\
 Target \vhunfi (\%) & 35.2     & 32.8 & 36.1  & 32.8      &  Target \vhunfi (\%) & 34.2     & 34.9 & 37.1  & 31.5 \\
 Target HI           & 0.63     & 0.65 & 0.63  & 0.64      &  Target HI           & 0.63     & 0.62 & 0.59  & 0.66 \\
                     &          &      &       &           &                      &          &      &       &         \\
Case 4 (V=80.2cc)    & Original & Cone & Cone2 & Firework  &  Case 9 (V=89.0cc)   & Original & Cone & Cone2 & Firework \\
 Bladder \vsevfi (cc)& 0.78     & 0.84 & 0.86  & 0.90      &  Bladder \vsevfi (cc)& 5.08     & 4.40 & 4.40  & 3.24 \\
 Rectum  \vsevfi (cc)& 0.35     & 0.76 & 0.87  & 0.48      &  Rectum  \vsevfi (cc)& 1.53     & 1.73 & 1.36  & 1.61 \\
 Urethra \vhund (cc) & 1.76     & 1.80 & 1.71  & 1.75      &  Urethra \vhund (cc) & 1.54     & 1.52 & 1.53  & 1.55 \\
 Urethra \vhuntw (cc)& 0.05     & 0.22 & 0.48  & 0.14      &  Urethra \vhuntw (cc)& 0.14     & 0.01 & 0.11  & 0.11 \\
 Target \vhund (\%)  & 90.9     & 92.7 & 90.2  & 90.3      &  Target \vhund (\%)  & 92.2     & 91.1 & 90.0  & 90.6 \\
 Target \vhunfi (\%) & 27.6     & 33.5 & 34.4  & 27.4      &  Target \vhunfi (\%) & 30.7     & 28.8 & 33.0  & 34.9 \\
 Target HI           & 0.70     & 0.64 & 0.62  & 0.70      &  Target HI           & 0.67     & 0.68 & 0.63  & 0.62 \\
                     &          &      &       &           &                      &          &      &       &         \\
Case 5 (V=43.4cc)    & Original & Cone & Cone2 & Firework  &  Case 10 (V=57.8cc)  & Original & Cone & Cone2 & Firework \\
 Bladder \vsevfi (cc)& 0.38     & 0.50 & 0.66  & 0.29      &  Bladder \vsevfi (cc)& 0.51     & 0.65 & 0.54  & 0.73 \\
 Rectum  \vsevfi (cc)& 0.34     & 0.23 & 0.28  & 0.16      &  Rectum  \vsevfi (cc)& 1.94     & 0.56 & 0.61  & 0.60 \\
 Urethra \vhund (cc) & 1.10     & 1.11 & 1.11  & 1.06      &  Urethra \vhund (cc) & 1.35     & 1.67 & 1.54  & 1.50 \\
 Urethra \vhuntw (cc)& 0.14     & 0.11 & 0.18  & 0.13      &  Urethra \vhuntw (cc)& 0.00     & 0.06 & 0.19  & 0.17 \\
 Target \vhund (\%)  & 96.8     & 95.6 & 95.9  & 94.1      &  Target \vhund (\%)  & 90.5     & 94.3 & 90.2  & 90.8 \\
 Target \vhunfi (\%) & 43.0     & 35.7 & 36.3  & 38.7      &  Target \vhunfi (\%) & 31.2     & 28.7 & 31.7  & 31.4 \\
 Target HI           & 0.56     & 0.63 & 0.62  & 0.59      &  Target HI           & 0.66     & 0.65 & 0.65  & 0.65 \\
 
    \dbline
   \end{tabular} 
  }  
 \end{center}
\end{table*}

\clearpage

\bibliographystyle{unsrt}
\bibliography{all.bib}

\end{document}